\documentclass[10pt,journal]{IEEEtran}
\usepackage{geometry}
\usepackage{verbatim}
\usepackage{amsmath,bm}
\usepackage{graphicx}
\usepackage{algorithm}
\usepackage{algorithmic}
\usepackage{booktabs}
\usepackage{multicol}
\usepackage{cite}
\usepackage{mathrsfs}
\usepackage{amsfonts,amssymb}
\usepackage{graphicx}
\usepackage{subfigure}
\usepackage{svg}

\usepackage[colorlinks,linkcolor=black,anchorcolor=black,citecolor=black]{hyperref}
\ifCLASSINFOpdf
\else
\fi
\begin{document}
%

\title{Distributed Cooperative Positioning in Dense Wireless Networks: A Neural Network Enhanced Fast Convergent Parametric Message Passing Method}

\author{Yue~Cao,
        Shaoshi~Yang,~\IEEEmembership{Senior Member,~IEEE},
        and Zhiyong~Feng,~\IEEEmembership{Senior Member,~IEEE}
\thanks{\textcolor{black}{This work was supported by the Beijing Municipal Natural Science Foundation (No. L242013 and Z220004). \textit{Corresponding author: Shaoshi Yang.}}}
\thanks{Y. Cao, S. Yang and Z. Feng are with the School of Information and Communication Engineering, Beijing University of Posts and Telecommunications, and the Key Laboratory of Universal Wireless Communications, Ministry of Education, Beijing, 100876, China (E-mails: \{caoyue, shaoshi.yang, fengzy\}@bupt.edu.cn).}

}

\pagestyle{empty} 
\maketitle
\thispagestyle{empty}  

\begin{abstract}
Parametric message passing (MP) is a promising technique that provides reliable marginal probability distributions for distributed cooperative positioning (DCP) based on factor graphs (FG), while maintaining minimal computational complexity. However, conventional parametric MP-based DCP methods may fail to converge in dense wireless networks due to numerous short loops on FG. Additionally, the use of inappropriate message approximation techniques can lead to increased sensitivity to initial values and significantly slower convergence rates. To address the challenging DCP problem modeled by a loopy FG, we propose an effective graph neural network enhanced fast convergent parametric MP (GNN-FCPMP) method. We first employ Chebyshev polynomials to approximate the nonlinear terms present in the FG-based spatio-temporal messages. This technique facilitates the derivation of globally precise, closed-form representations for each message transmitted across the FG, and reduces MP's sensitivity to initial positional values. Then, the parametric representations of spatial messages are meticulously refined through data-driven GNNs. Conclusively, by performing inference on the FG, we derive more accurate closed-form expressions for the \textit{a posteriori} distributions of node positions. Numerical results substantiate the capability of GNN-FCPMP to significantly enhance positioning accuracy within wireless networks characterized by high-density loops and ensure rapid convergence.
\end{abstract}


\IEEEpeerreviewmaketitle

\section{Introduction}
\IEEEPARstart{A}{ccurate} position sensing is crucial for a myriad of emerging vertical applications in future communications, serving as a foundational element that enhances functionality and efficiency across diverse sectors \cite{Wymeersch2009}. The distributed cooperative positioning (DCP) \cite{Wymeersch2009, Lv_2016, cao2023spatial, cao2022geo} technology, which has higher flexibility, scalability and robustness than centralized implementations, is pivotal in advancing modern sensing systems in environments lacking global satellite navigation support. Within the DCP paradigm, agents (nodes whose positions are unknown) enhance their position accuracy and reliability by collective cooperative spatial and temporal data integration from both external and internal measurements with other agents.

As one of the most classical approaches, factor graph (FG) based message passing (MP) is exceptionally equipped to furnish nodes with precise \textcolor{black}{posterior} probability distributions in the singly connected graph \cite{weiss1999correctness,NEBP2021}, by locally marginalizing the joint \textit{a posteriori} distribution. MP typically involves two phases: message representation and message multiplication. \textcolor{black}{The former abstracts cooperation information across temporal and spatial dimensions,} while the latter updates the belief about agent positions \textcolor{black}{and} the \textit{a posteriori} probability distributions. In the context of DCP that employs FG-based MP, the inherent non-linearity of ranging models necessitates the transmission of approximate messages between nodes. The predominant methods for representing these approximate messages are categorized into two types: particle-based \cite{SPAWN-comparison,NEBP-CL2021} and parametric methods \cite{SPAWN-comparison,SPATE,cao2022geo,cao2023cl}. Specifically, particle-based methods employ thousands of weighted particles for detailed message depiction, ensuring high fidelity but significantly increasing computational demands on agents \cite{Wymeersch2009,SPAWN-comparison}. \textcolor{black}{In contrast, parametric methods model each message with specific parameters, such as a two-dimensional toroidal distribution with six parameters in 2D scenarios \cite{SPAWN-comparison}. However, the approach in \cite{SPAWN-comparison}} lacks a closed-form solution for the agent's \textit{a posteriori} distribution, requiring numerical methods for approximation. Alternatively, the authors in \cite{SPATE,cao2023cl} use one-dimensional Gaussian distributions for messages, which may bias positioning due to inaccurate assumptions. Additionally, employing Taylor polynomials for nonlinear terms \cite{SPATE,cao2023cl} leads to high sensitivity to the \textit{a priori} distribution, due to their limited approximation capacity away from the expansion point, often slowing convergence.

In wireless networks with a high density of agents, the FG or other probabilistic graphical models describing these networks typically contain many loops. Such loops introduce inaccuracies in MP inference and complicate convergence, which significantly reduces both positioning accuracy and convergence speed in DCP \cite{NEBP2021,NEBP-CL2021,kirkley2021belief,cantwell2019message}. Some advancements have been made to adapt MP for use in graph models that include loops. Notably, the authors of \cite{cantwell2019message} developed an MP framework that addresses the bond percolation problem in looped networks and calculates spectra for large sparse symmetric matrices by incorporating loops into the MP equations. Building on this, the authors of \cite{kirkley2021belief} expanded the methodology for general probabilistic models, enhancing the utility of the \cite{cantwell2019message} approach. While this method provides precise results on short loops that do not exceed a certain length, it only yields approximate outcomes for longer, arbitrary loops. Moreover, its implementation demands detailed knowledge of the graph’s global structure and incurs exponentially higher computational costs with increasing loop size. \textcolor{black}{Recent years have seen growing interests in data-driven deep learning techniques to enhance MP in graph models \cite{MPNN2017,NEBP2021,NEBP-CL2021,MPNN-LSTM}.} Graph neural network (GNN) stands out as a connectionist model adept at capturing node dependencies. \textcolor{black}{To combine model-based inference and data-driven neural networks for molecular property prediction in quantum chemistry, \cite{MPNN2017} proposed the MP neural network (MPNN) framework, abstracting commonalities among existing neural networks for graph-structure data.} The study in \cite{NEBP2021} developed a neural enhanced belief propagation (NEBP) model that leverages a trained GNN to refine the original messages passed on the FG. 

However, the methods from the aforementioned studies are not readily applicable to DCP due to their reliance on extensive global graph structure knowledge, which is generally unavailable to individual agents in distributed wireless networks. Recently, the NEBP model was adapted to tackle the DCP challenge, as noted in \cite{NEBP-CL2021}. However, this method is marred by high complexity due to the extensive use of particles and iterative message computations. Furthermore, it fails to consider the critical temporal information from hardware measurements, which forms an essential part of the \textit{a priori} knowledge for nodes in FG. Most recently, the authors of \cite{MPNN-LSTM} introduced a data-driven approach by integrating Long Short-Term Memory (LSTM) modules with MPNN. This model facilitates DCP through a message-passing-like mechanism, achieving acceptable efficacy when the number of cooperative nodes is large. Despite its potential, the MPNN-LSTM \cite{MPNN-LSTM} still struggles with reduced positioning accuracy in networks with a high density of loops.

In light of the challenges highlighted previously, this paper introduces a novel approach, dubbed GNN enhanced fast convergent parametric MP (GNN-FCPMP), that combines GNN with FG based MP to address the DCP problem in dense wireless networks. Our GNN-FCPMP aims to provide high-accuracy closed-form representations of messages and the \textit{a posteriori} distribution of agent's position.

The key contributions of our study are outlined below:
\begin{itemize}

\item[$\bullet$] Different from \cite{SPATE,cao2022geo,cao2023spatial,li2019convergence}, our approach relies on the well-founded assumption that spatio-temporal messages can be approximated either as a single two-dimensional toroidal distribution or the superposition of two toroidal distributions \cite{Wymeersch2009,SPAWN-comparison}, which is more congruent with the ranging model. In our high-precision parametric message representation method, we derive the closed-form representations for each message passed on FG by exploiting the Chebyshev polynomials to approximate the nonlinear terms in the messages. Notably, this includes messages from agents with ambiguous positions (the distribution of agent's position is bimodal toroidal distribution), a feat that other parametric message representation methods have not fulfilled. Our method employs a global approximation approach, which significantly reduces the dependency of agents on their initial estimated positions. In contrast, the method described in \cite{SPATE,cao2022geo,cao2023spatial} is sensitive to the initial positions of the nodes.

\item[$\bullet$] To the best of our knowledge, the work in this paper constitutes the first effort to introduce the GNN into a parametric MP method for solving the challenging DCP problem modeled by loopy FG. Building upon our high-accuracy parametric message representation, the proposed GNN-FCPMP enjoys a significantly reduced computational complexity of representing both the spatio-temporal messages and the node embedding on the corresponding FG-based subgraph. More fundamentally, our GNN-FCPMP is capable of obtaining closed-form expressions of the \textit{a posteriori} distributions, thus dramatically reducing the computational complexity of \textit{message representation} and \textit{message multiplication} \cite{SPAWN-comparison} involved in a wide range of applications. In addition, the proposed GNN-FCPMP method also has a higher positioning accuracy than the contribution of \cite{SPAWN-comparison,cao2023cl,NEBP-CL2021,MPNN-LSTM}.

\end{itemize}
\section{System Model and Problem Formulation}
Consider a wireless network comprising $N$ agents and $A$ anchors, with the positions of the anchors known \textit{a priori}. The network operates on a slotted time schedule. We denote the set of anchors and the specific agents that send signals to agent $i$ during time slot $t$ as $\mathbb{A}_{\rightarrow i}^{t}$ and $\mathbb{U}_{\rightarrow i}^{t}$, respectively. The position vector of agent $i$ at time slot $t$ is defined as $\bm{x}_{i}^{t} \triangleq [x_{i}^{t}, y_{i}^{t}]^{\text{T}}$, with $(\cdot)^{\text{T}}$ representing the transpose operation.

At each time slot $t$, agent $i$ collects both external measurements from its neighbors and internal measurements. The noise-contaminated external ranging measurement\footnote{All the links considered are line-of-sight (LOS), while the non-line-of-sight (NLOS)/LOS mixed environment can be considered in a similar manner presented in our previous work \cite{cao2022geo}.} from node $j$ (either an agent or an anchor) to agent $i$ satisfies $z_{j \rightarrow i}^t=d_{ij}^t+e_{j \rightarrow i}^t$, where $d_{ij}^t$ is the Euclidean distance between node $j$ and agent $i$ at time slot $t$, and $e_{j \rightarrow i}^t \sim \mathcal{N}(0, (\sigma_{j \rightarrow i}^t)^2)$ denotes the measurement error, modeled as a zero-mean Gaussian distribution with variance $(\sigma_{j \rightarrow i}^t)^2$. Similarly, the internal measurement $z^t_{i,\text{int}}$ also follows a Gaussian distribution with zero-mean and variance $(\sigma_{i,\text{int}}^t)^2$. We aggregate all noisy ranging measurements (both external and internal) received by agent $i$ at time slot $t$ into the vector \textcolor{black}{$\bm{z}_i^t = [z^t_{i,\text{int}}, z_{j \rightarrow i}^t, \dots], j \in \mathbb{A}_{\rightarrow i}^{t} \cup \mathbb{U}_{\rightarrow i}^{t}$.} Our goal is to infer the \textit{a posteriori} probability distribution of agent $i$’s position at any time slot $t$, based solely on these noisy measurements from that time slot, i.e., $p(\bm{x}_i^t | \bm{z}_i^t)$.

\section{The Proposed GNN-FCPMP Approach}

\subsection{Message Representation}
\begin{figure}[t]
\begin{center}
\includegraphics[scale=0.26]{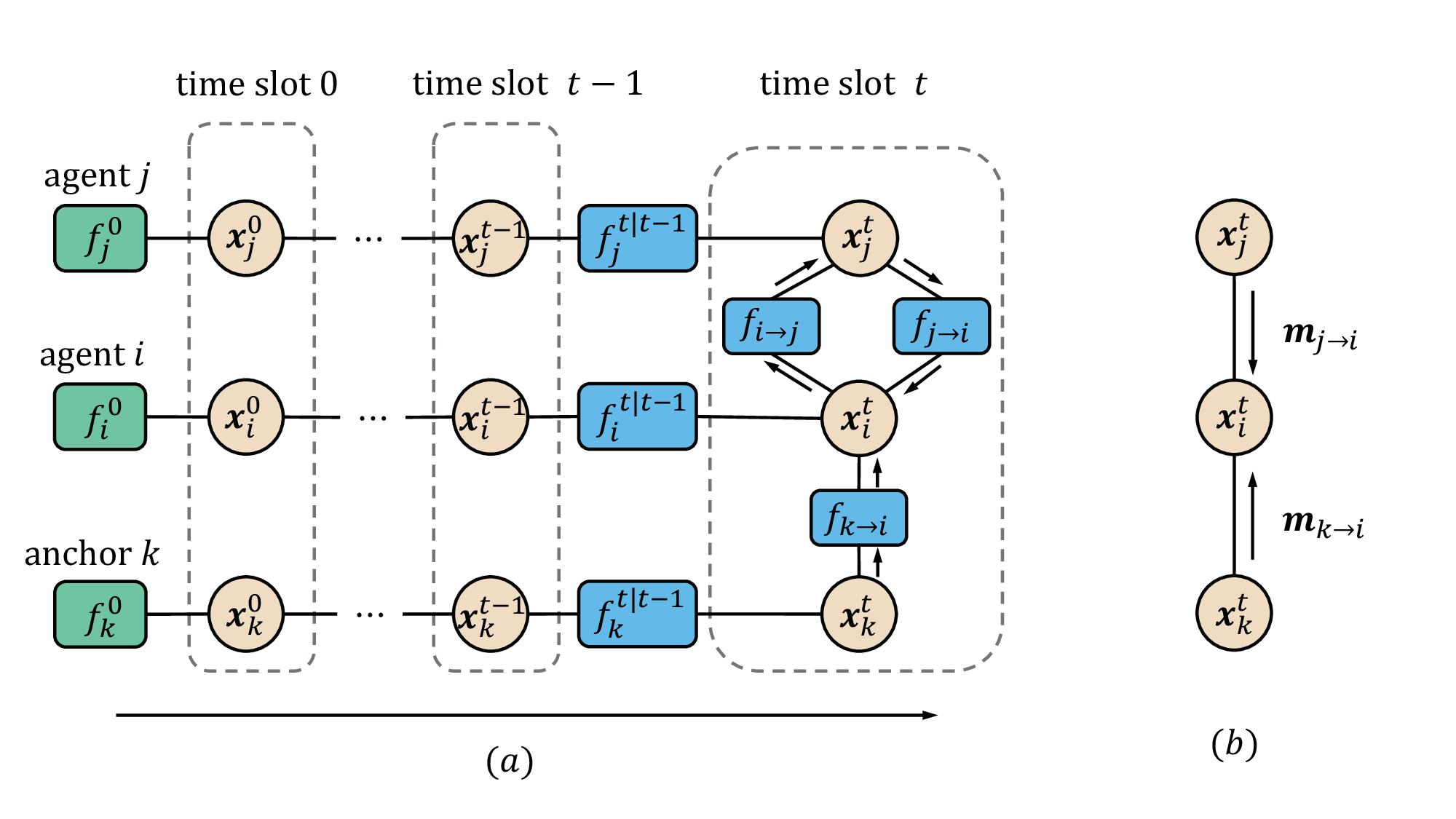}
\end{center}
\vspace*{-5mm}
\caption{(a) FG of $p(\bm{x}_i^{t} | \bm{z}_i^t)$, where we have node \(j \in \mathbb{U}_{\rightarrow i}^{t}\), node \(k \in \mathbb{A}_{\rightarrow i}^{t}\), \(f_i^0=p\left(\bm{x}_i^0\right)\), \(f_i^{t|t-1}= p\left(\bm{x}_i^{t} \mid \bm{x}_i^{t-1}\right) p\left(z_{i, \text {int}}^{t} \mid \bm{x}_i^{t-1}, \bm{x}_i^{t}\right)\), and \(f_{j \rightarrow i}=p\left(z_{j \rightarrow i}^{t} \mid \bm{x}_{i}^{t}, \bm{x}_{j}^{t}\right)\). (b) The equivalent graph representation in GNN (referred to as \textit{GNN graph} in what follows) for the FG-based subgraph of agent $i$.}
\label{fig:1} 
\vspace*{-5mm}
\end{figure}

In MP, the message representation of probabilistic information for transmission and computation significantly influences the complexity and effectiveness of DCP algorithm. To establish the parametric representation of each message on the FG, we initially construct the FG by decomposing the \textit{a posteriori} distribution of position-vector estimates. This decomposition maps both spatial and temporal operations of nodes onto the FG, thereby supporting the distributed implementation of wireless DCP. Consequently, the marginal distribution of the position for agent $i$ is factorized as follows:
\begin{equation}
\begin{aligned} p(\bm{x}_i^t | \bm{z}_i^t)  = & p(\bm{x}_i^t | \bm{z}_i^t, \bm{x}_i^{t-1}, \bm{x}_j^t)
\\ \propto  & p\left(\bm{x}_i^{t-1}|\bm{z}_i^{t-1}\right) p\left(z_{i,\text{int}}^t | \bm{x}_i^t, \bm{x}_i^{t-1}\right) \\ 
 &\times p\left(\bm{x}_i^t | \bm{x}_i^{t-1}\right) \prod_{j \in \mathbb{A}_{\rightarrow i}^t \cup \mathbb{U}_{\rightarrow i}^t}
p\left(z_{j \rightarrow i}^t | \bm{x}_i^t, \bm{x}_j^t\right). \end{aligned}
\end{equation}
The FG of \(p(\bm{x}_i^t | \bm{z}_i^t)\) has a structure illustrated in Fig. \ref{fig:1}(a). Then we exploit MP to infer the \textit{a posteriori} distribution of agent $i$'s position. According to MP, the \textit{a posteriori} distribution of $\bm{x}_i^t$, i.e., $p\left(\bm{x}_i^t \mid \bm{z}_i^t\right)$ satisfies
\begin{equation}
\begin{aligned}
p\left(\bm{x}_i^t \mid \bm{z}_i^t\right)& \approx b_{\ell_\text{max}}(\bm{x}_i^t)\\
& \propto \mu_{f_i^{t|t-1}\rightarrow \bm{x}_i^t} \prod_{j \in \mathbb{U}_i^t \cup \mathbb{A}_i^t} \mu_{\ell_\text{max}, f_{j \rightarrow i} \rightarrow \bm{x}_i^t}, \label{a_posteriori}
\end{aligned}
\end{equation}
where $b_{\ell_\text{max}}(\bm{x}_i^t)$ denotes the belief of $\bm{x}_i^t$ at iteration $\ell_\text{max}$, $\ell_\text{max}$ denotes the maximum number of iterations, $\mu_{f_i^{t|t-1}\rightarrow \bm{x}_i^t}$ is the temporal message passed from factor $f_i^{t|t-1}$ to variable $\bm{x}_i^t$, $\mu_{\ell_\text{max}, f_{j \rightarrow i} \rightarrow \bm{x}_i^t}$ represents the spatial message passed from factor $f_{j \rightarrow i}$ to variable $\bm{x}_i^t$ at iteration $\ell_\text{max}$. Since the proposed GNN-FCPMP is fully distributed, let us examine the messages received by agent $i$ at time slot $t$ and iteration $l$ as illustrative examples. The temporal message that is passed from factor $f_i^{t|t-1}$ to variable $\bm{x}_i^t$ conforms to the following criteria\footnote{MP can operate on unnormalized beliefs \cite{Wymeersch2009,NEBP2021}.}:
\begin{equation}
\mu_{f_i^{t|t-1} \rightarrow \bm{x}_i^t} \propto \exp \left\{-\frac{\left(z^t_{i, \text{int}}-\xi_1\right)^2}{2 \sigma_{i, \text{int}}^2}\right\}, \label{temporal_message_precise}
\end{equation}
where
\begin{equation}
\xi_1= \|\bm{x}_i^t-\hat{\bm{x}}_i^{t-1}\|_2 = \sqrt{\left(x_i^t-\hat{x}_i^{t-1}\right)^2+\left(y_i^t-\hat{y}_i^{t-1}\right)^2},
\end{equation}
$\|\cdot\|_2$ represents the Euclidean norm, and $\hat{\bm{x}}_i^{t-1} = \left(\hat{x}_i^{t-1}, \hat{y}_i^{t-1}\right)$ is the estimated value of $\bm{x}_i^{t-1}$. Specifically, the temporal message exhibits a distribution that is akin to a toroidal (donut-like) form. It peaks along a circular path where the Euclidean distance from the center $\left(\hat{x}_i^{t-1}, \hat{y}_i^{t-1}\right)$ equals $z^t_{i, \text{int}}$. This configuration creates a visual and mathematical ``donut" shape, characterized by a central ring of high probability density that smoothly tapers off following a Gaussian decay both towards the center and outward. To enhance the efficiency of message representation and reduce the computational complexity faced by agents, $\xi_1$ is approximated using second-order Chebyshev polynomials\footnote{\textcolor{black}{Due to Runge's phenomenon, higher-order Chebyshev polynomials do not improve accuracy significantly, making second-order polynomials a better choice for resource-constrained agents considering the computational complexity.}}, satisfying
\begin{equation}
\xi_1 \approx \sum_{n=0}^2 \sum_{m=0}^2 c_{n m} T_n(x) T_m(y), \label{xi_approx}
\end{equation}
where $c_{n m}$ denotes the coefficient of the Chebyshev polynomials, $T_n(x)$ and $T_m(y)$ represent the polynomial basis of $x_i^t$ and $y_i^t$, respectively, satisfy
\begin{equation}
\begin{aligned}
& T_0(x)=1, T_1(x)=x, \\
& T_{n+1}(x)=2 x T_n(x)-T_{n-1}(x) . \label{T_n_x_y}
\end{aligned}
\end{equation}
And the matrix $\bm{C}$, which includes all the coefficients $c_{n m}$ of the third-order polynomials, can be calculated using specialized mathematical software. This matrix satisfies
\begin{equation}
\begin{aligned}
\bm{C} = \begin{bmatrix}
    c_{00} & c_{01} & c_{02} \\
    c_{10} & c_{11} & c_{12} \\
    c_{20} & c_{21} & c_{22} \\
\end{bmatrix}
= \begin{bmatrix}
    1.12  & -0.45 & 0.20 \\  
    -0.45 & -0.14 & 0.06 \\
    0.20  & 0.06 & -0.02 \\
\end{bmatrix}. \label{Matrix_C}
\end{aligned}
\end{equation}

Upon substituting \eqref{xi_approx}, \eqref{T_n_x_y}, and \eqref{Matrix_C} into \eqref{temporal_message_precise}, the temporal message $\mu_{f_i^{t|t-1} \rightarrow \bm{x}_i^t}$ is given by
\begin{equation}
\mu_{f_i^{t|t-1} \rightarrow \bm{x}_i^t} \propto \exp  \left\{\bm{\omega}_i^\mathrm{T} \bm{\psi} \right\}, \label{temporal_message_closed_form}
\end{equation}
where 
\begin{equation}
\bm{\omega}_i = [\omega_{i1}, \omega_{i2}, \omega_{i3}, \omega_{i4}, \omega_{i5}]^\mathrm{T}, \label{omega_i}
\end{equation}
\begin{equation}
\bm{\psi} = [-(x_i^t)^2, -(y_i^t)^2, x_i^t,  y_i^t, x_i^t y_i^t]^\mathrm{T},
\end{equation}
and
\begin{equation}
\left\{\begin{array}{l}
\omega_{i1}=\frac{-4z^t_{i, \text{int}}c_{20}+4z^t_{i, \text{int}}c_{22}+1}{2 (\sigma_{i,\text{int}}^t)^2}, \\
\omega_{i2}=\frac{-4z^t_{i, \text{int}}c_{02}+4z^t_{i, \text{int}}c_{22}+1}{2 (\sigma_{i,\text{int}}^t)^2}, \\
\omega_{i3}=\frac{z^t_{i, \text{int}}c_{10}-z^t_{i, \text{int}}c_{12}+\hat{x}_i^{t-1}}{(\sigma_{i,\text{int}}^t)^2}, \\
\omega_{i4}=\frac{z^t_{i, \text{int}}c_{01}-z^t_{i, \text{int}}c_{21}+\hat{y}_i^{t-1}}{(\sigma_{i,\text{int}}^t)^2}, \\
\omega_{i5}=\frac{z^t_{i, \text{int}} c_{11}}{(\sigma_{i,\text{int}}^t)^2}. \\ 
\end{array}\right.
\end{equation}

Similar to the processing of temporal message, the spatial message passed from anchor $k \in \mathbb{A}_{\rightarrow i}^t$ to agent $i$ at time slot $t$ and iteration $\ell_\text{max}$ is given by
\begin{equation}
\mu_{\ell_\text{max}, f_{k \rightarrow i} \rightarrow \bm{x}_i^t}\propto \exp  \left\{\bm{\omega}_{k}^\mathrm{T} \bm{\psi} \right\}, \label{spatial_message_anchor_closed_form}
\end{equation}
where
\begin{equation}
\bm{\omega}_{k} = [\omega_{k1}, \omega_{k2}, \omega_{k3}, \omega_{k4}, \omega_{k5}]^\mathrm{T}, \label{omega_k}
\end{equation}
$\bm{\omega}_{k}$ is similar to $\bm{\omega}_{i}$ and is omitted here.

As for the spatial message passed from agent $j \in \mathbb{U}_{\rightarrow i}^t$ to agent $i$, when the distribution of agent $j$'s position is bimodal \cite{Wymeersch2009}, it satisfies
\begin{equation}
\begin{aligned}
\mu_{\ell, f_{j\rightarrow i} \rightarrow \bm{x}_i^t} \propto
& \exp \left\{-\frac{\left(z^t_{j\rightarrow i}-\|\bm{x}_i^t - \bm{x}_1\|\right)^2}{2 \sigma_{j\rightarrow i}^2}\right\} \\
&+ \exp \left\{-\frac{\left(z^t_{j\rightarrow i}-\|\bm{x}_i^t - \bm{x}_2\|\right)^2}{2 \sigma_{j\rightarrow i}^2}\right\}, \label{spatial_message_j2i_precise}
\end{aligned}
\end{equation}
where $\bm{x}_1 = \left[x_1, y_1\right]$ and $\bm{x}_2 = \left[x_2, y_2\right]$ are the two midpoints of the bimodal distribution. Similarly, $\|\bm{x}_i^t - \bm{x}_1\|$ and $\|\bm{x}_i^t - \bm{x}_2\|$ are approximated by invoking the second-order Chebyshev polynomials, yielding
\begin{equation}
\mu_{f_{j \rightarrow i} \rightarrow \bm{x}_i^t} \propto \exp  \left\{\bm{\omega}_j^\mathrm{T} \bm{\psi} \right\}, \label{spatial_message_agent_closed_form}
\end{equation}
where 
\begin{equation}
\bm{\omega}_j = [\omega_{j1}, \omega_{j2}, \omega_{j3}, \omega_{j4}, \omega_{j5}]^\mathrm{T}, \label{omega_j}
\end{equation}
and

\begin{equation}
\left\{\begin{array}{l}
\omega_{j1}=\frac{-4z^t_{j \rightarrow i}c_{20}+4z^t_{j \rightarrow i}c_{22}+1}{(\sigma_{j \rightarrow i}^t)^2}, \\
\omega_{i2}=\frac{-4z^t_{j \rightarrow i}c_{02}+4z^t_{j \rightarrow i}c_{22}+1}{(\sigma_{j \rightarrow i}^t)^2}, \\
\omega_{i3}=\frac{2z^t_{j \rightarrow i}c_{10}-2z^t_{j \rightarrow i}c_{12}+x_1+x_2}{(\sigma_{j \rightarrow i}^t)^2}, \\
\omega_{i4}=\frac{2z^t_{j \rightarrow i}c_{01}-2z^t_{j \rightarrow i}c_{21}+y_1+y_2}{(\sigma_{j \rightarrow i}^t)^2}, \\
\omega_{i5}=\frac{2z^t_{j \rightarrow i} c_{11}}{(\sigma_{j \rightarrow i}^t)^2}. \\ 
\end{array}\right.
\end{equation}
When the distribution of agent $j$'s position is unimodal, the parametric representation of $\mu_{f_{j \rightarrow i} \rightarrow \bm{x}_i^t}$ is similar to $\mu_{f_{k \rightarrow i} \rightarrow \bm{x}_i^t}$ and is omitted here.

We consider the computational complexity of a single agent in a single iteration since this paper focuses on distributed schemes. We use $N_\text{s}$ and $N_\text{rel}$ to denote the number of particles and neighbors, respectively. The computational complexity of our parametric message representation and particle-based representation \cite{SPAWN-comparison,NEBP-CL2021} are $\mathcal{O}(N_\text{rel})$ and $\mathcal{O}(N_\text{s}^2 N_\text{rel})$, respectively.

\subsection{Message Enhancement}
\begin{figure}[t]
\begin{center}
\includegraphics[scale=0.26]{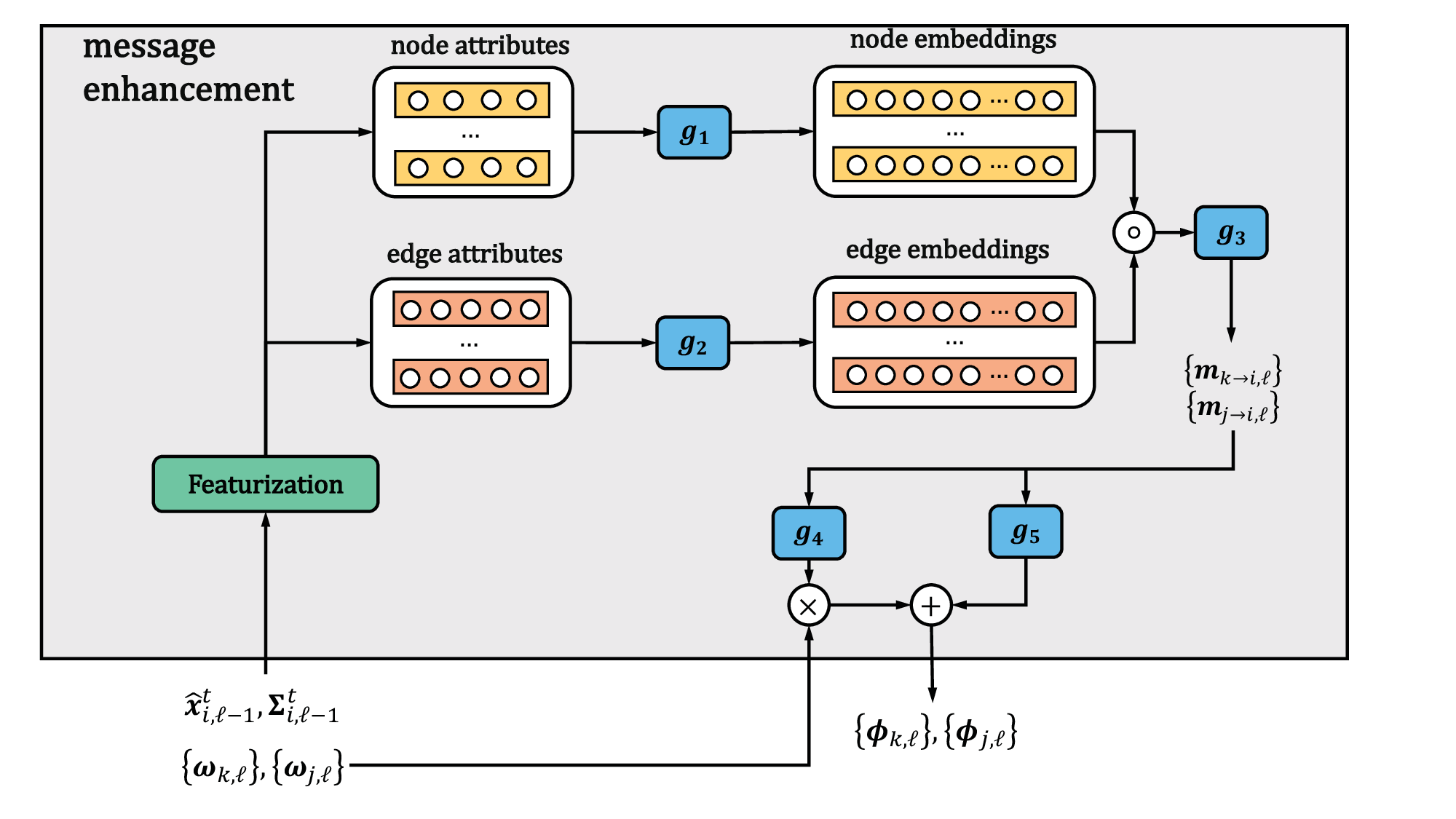}
\end{center}
\vspace*{-5mm}
\caption{The illustration of the message enhancement stage.}
\label{fig:2} 
\vspace*{-5mm}
\end{figure}
The message enhancement stage builds on the foundation of the message representation, aiming to optimize the efficiency of the MP process. The \textit{GNN graph} corresponding to the FG-based subgraph of agent $i$ at time slot $t$ is shown in Fig. \ref{fig:1}(b), where \(\bm{m}_{j \rightarrow i}\) represents the message passed on the corresponding graph model from vertex $\bm{x}_j^t$ to vertex $\bm{x}_i^t$, and its value is yet to be obtained by learning. 

Corresponding to its local FG, agent $i$ creates the node embedding vectors for each node on the GNN graph by
\begin{equation}
\bm{h}_{i,\ell} = g_1\left(\bm{a}_{i,\ell}\right),
\end{equation}
where $\bm{a}_{i,\ell} = [\hat{x}_{i, \ell-1}^t,\hat{y}_{i, \ell-1}^t,\sigma^2_{x_i^t, \ell-1},\sigma^2_{y_i^t, \ell-1}]$ denotes the node attribute vector. And the edge embedding vectors for each edge on GNN graph satisfies
\begin{equation}
\bm{h}_{j \rightarrow i,\ell} = g_2\left(\bm{a}_{j \rightarrow i,\ell}\right),
\end{equation}
where $\bm{a}_{j \rightarrow i,\ell}=\bm{\omega}_{j,\ell}$ denotes the edge attribute vector of dimension 5. The GNN-generated messages sent from node $j$ to agent $i$ is given by
\begin{equation}
\bm{m}_{j \rightarrow i,\ell} = g_3\left(\bm{h}_{j \rightarrow i,\ell} \circ \bm{h}_{j,\ell}\right),  \label{MPGNN}
\end{equation}
where $\circ$ denotes the element-wise multiplication. Then, agent $i$ exploits $\{\bm{m}_{j \rightarrow i, \ell}\}$ to refine the spatial message representations. The refined parametric representations are given by
\begin{equation}
\begin{aligned}
\bm{\phi}_{j, \ell}=&[\phi_{j1}, \phi_{j2}, \phi_{j3}, \phi_{j4}, \phi_{j5}]^\mathrm{T} \\
=&\bm{\omega}_{j, \ell} g_4\left(\bm{m}_{j \rightarrow i, \ell}\right)+g_5\left(\bm{m}_{j \rightarrow i, \ell}\right), \label{GNN_refined_spatio_message_representation}   
\end{aligned}
\end{equation}
where $g_4(\cdot)$ and $g_5(\cdot)$ are GNNs with trainable parameters. Specifically, $g_4(\cdot)$ outputs a positive scalar and $g_5(\cdot)$ outputs a positive vector of dimension 5. The corresponding refined spatial message satisfies
\begin{equation}
\zeta_{\ell,f_{j \rightarrow i} \rightarrow x_i^t} \propto \exp  \left\{\bm{\phi}_{j, \ell}^\mathrm{T} \bm{\psi} \right\}.\label{GNN_refined_spatio_message}
\end{equation}
In conclusion, the illustration of the message enhancement stage is shown in Fig. \ref{fig:2}.

\subsection{Message Multiplication}
In message multiplication, agent $i$ computes and updates its belief about its own position during time slot $t$ by fusing temporal information from itself with spatial information from its neighbors. At iteration $\ell_\text{max}$, upon substituting \eqref{temporal_message_closed_form} and \eqref{GNN_refined_spatio_message} into \eqref{a_posteriori}, the \textit{a posteriori} distribution of agent $i$'s position at time slot $t$ satisfies
\begin{equation}
p(\bm{x}_i^t|\bm{z}_i^t) \propto \exp \left\{-\frac{1}{2}(\bm{x}_i^t-\bm{m}_{\bm{x}_i^t|\bm{z}_i^t})^\mathrm{T} \bm{\Sigma}_{\bm{x}_i^t|\bm{z}_i^t}^{-1}(\bm{x}_i^t-\bm{m}_{\bm{x}_i^t|\bm{z}_i^t})\right\},\label{GNN_refined_a_posteriori}
\end{equation}
where
\begin{equation}
\bm{m}_{\bm{x}_i^t|\bm{z}_i^t} = {\left[\frac{\sum \phi_{j3} + \omega_{i3}}{2\left(\sum \phi_{j1} + \omega_{i1} \right)} ,\right.} \left. \frac{\sum \phi_{j4} + \omega_{i4}}{2\left(\sum \phi_{j2} + \omega_{i2} \right)}\right],
\end{equation}
\begin{equation}
\bm{\Sigma}_{\bm{x}_i^t \mid \bm{z}_i^t}=\left[\begin{array}{cc}
\sigma_1^2 & \sigma_2^2 \\
\sigma_2^2 & \sigma_3^2
\end{array}\right],
\end{equation}
and $\sigma_1^2 = \left(\sum \phi_{j1} + \omega_{i1} \right)^{-1}$, $\sigma_2^2 = \frac{1}{2}\left(\sum \phi_{j5} + \omega_{i5} \right)$, and $\sigma_3^2 = \left(\sum \phi_{j2} + \omega_{i2}\right)^{-1}$, respectively, where $j \in \mathbb{U}_i^t \cup \mathbb{A}_i^t$. At any time slot $t$, each agent can determine the minimum mean squared error (MMSE) estimate of its own position by taking the mean value involved in $p(\bm{x}_i^t|\bm{z}_i^t)$ \cite{Wymeersch2009,SPAWN-comparison}. The above step is repeated recursively for $\ell_\text{max}$ iterations.

\section{Experiment Results and Discussions}

\subsection{Dataset and Training Settings}

\begin{table*}[t]
\vspace*{5mm}
\centering
\caption{Details of the GNN structures}
\vspace*{-1mm}
\begin{tabular}{|cc|cc|cc|cc|cc|}
\hline
  \multicolumn{2}{|c|}{$g_1(\cdot )$} & \multicolumn{2}{c|}{$g_2(\cdot )$} & \multicolumn{2}{c|}{$g_3(\cdot )$} & \multicolumn{2}{c|}{$g_4(\cdot )$} & \multicolumn{2}{c|}{$g_5(\cdot )$} \\ \hline
  \multicolumn{1}{|c|}{Input} & $4 \times 1$ & \multicolumn{1}{c|}{Input} & $5 \times 1$ & \multicolumn{1}{c|}{Input} & $8 \times 1$ &\multicolumn{1}{|c|}{Input} & $8 \times 1$ & \multicolumn{1}{|c|}{Input} & $8 \times 1$ \\
  \multicolumn{1}{|c|}{Linear+ReLU} & $32 \times 1$ & \multicolumn{1}{c|}{Linear+ReLU} & $32 \times 1$ & \multicolumn{1}{c|}{Linear+ReLU} & $32 \times 1$ & \multicolumn{1}{c|}{Linear+ReLU} & $8 \times 1$ & \multicolumn{1}{c|}{Linear+ReLU} & $8 \times 1$ \\
  \multicolumn{1}{|c|}{Linear+ReLU} & $16 \times 1$ & \multicolumn{1}{c|}{Linear+ReLU} & $16 \times 1$ & \multicolumn{1}{c|}{Linear+ReLU} & $16 \times 1$ & \multicolumn{1}{c|}{Linear} & $1 \times 1$ & \multicolumn{1}{c|}{Linear} & $5 \times 1$ \\
  \multicolumn{1}{|c|}{Linear+ReLU} & $8 \times 1$ & \multicolumn{1}{c|}{Linear+ReLU} & $8 \times 1$ & \multicolumn{1}{c|}{Linear+ReLU} & $8 \times 1$ & \multicolumn{1}{c|}{} & & \multicolumn{1}{c|}{} & \\ \hline
\end{tabular}
\label{tab1}
\vspace*{-2mm}
\end{table*}  

Consider a wireless network covering a $[0, 200] \text{m} \times [0, 200] \text{m}$ area, with 50 mobile agents and 13 static anchors initially distributed within a $[20, 180] \text{m} \times [20, 180] \text{m}$ zone. Agents feature a 20m communication radius and a 20 iteration maximum. Noise in the ranging measurements is modeled with a variance of 0.1. Each agent moves a distance $d_i^t \sim \mathcal{N}(2,1)$ in a direction $\theta_i^t \sim \mathcal{U}(0,2\pi)$. To maintain a stable agent population, a new agent is placed into the area whenever an existing agent exits. We collected 1000 position realizations over 40 time slots as training data, assessing losses via root mean square error (RMSE) between estimated and actual positions. Optimization is conducted using the Adam optimizer, starting with a learning rate of \(2 \times 10^{-3}\) and decreasing to \(10^{-5}\). The details of GNN structures are shown in Table \ref{tab1}.

\subsection{Results and Discussions}
To assess the generalization capabilities of our GNN-FCPMP, we evaluated it in networks of varying sizes. This setup included 41 static anchors in a $[0,500]$m $\times$ $[0,500]$m area and 50 to 150 agents uniformly distributed within a $[50,450]$m $\times$ $[50,450]$m region, with a communication radius of 50m. We focused on triangular loops formed by variable nodes representing agent positions. In scenarios with 50, 100, and 150 agents, the average numbers of loops in the network, determined using the Monte Carlo method, were 23.74, 195.5, and 672.5, respectively.

\begin{figure}[t]
\vspace*{-5mm}
\begin{center}
\includegraphics[scale=0.42]{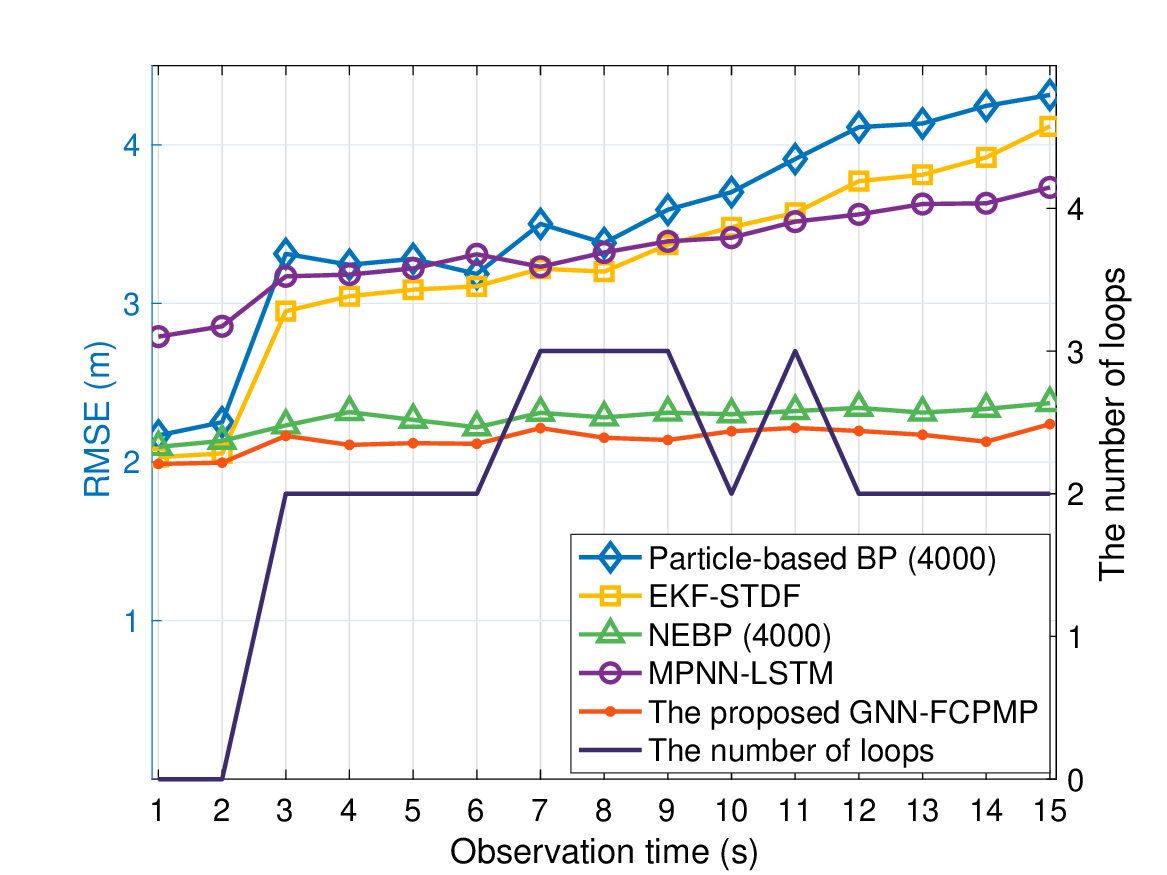}
\end{center}
\vspace*{-5mm}
\caption{RMSE performance of an agent with different number of loops, when using various DCP schemes.}
\label{Fig_3} 
\vspace*{-5mm}
\end{figure} 

We initially focus on a single agent to assess how varying loop numbers affect DCP performance. We set the agent count to 110, with $\ell_{\max} = 10$, and each agent moves a distance $d_i^t \sim \mathcal{N}(5,1)$ in a direction $\theta_i^t \sim \mathcal{U}(0, 2\pi)$. Fig.~\ref{Fig_3} compares the RMSE results of our GNN-FCPMP with EKF-STDF \cite{cao2023cl}, particle-based BP \cite{Wymeersch2009,SPAWN-comparison}, NEBP \cite{NEBP-CL2021}, and MPNN-LSTM \cite{MPNN-LSTM}. The curves labeled \textcolor{black}{``particle-based BP (4000)" and ``NEBP (4000)"} show their performance with 4000 particles. The x-axis numbers indicate the loop count during each observation period. For example, the agent is not involved in any loops during the first two time slots and is within two loops from time slots 3 to 6.

Our observations from Fig.~\ref{Fig_3} are twofold. Initially, we note that the accuracy of the particle-based BP, MPNN-LSTM, and our EKF-STDF significantly declines when the agent is within multiple loops, particularly evident after the third time slot. This substantial performance disparity between our GNN-FCPMP and EKF-STDF highlights the cumulative effect of positioning errors when the agent remains within loops for extended periods. Additionally, our GNN-FCPMP and the existing NEBP model demonstrate superior robustness, effectively mitigating performance degradation despite an increasing number of loops. This robustness confirms that the GNN's data-driven approach significantly refines message accuracy, thereby enhancing overall positioning outcomes. Notably, our GNN-FCPMP surpasses the performance of NEBP with 4000 particles, largely due to its precise parametric message representations and effective use of temporal information.

\begin{figure}[t]
\vspace*{-5mm}
\begin{center}
\includegraphics[scale=0.42]{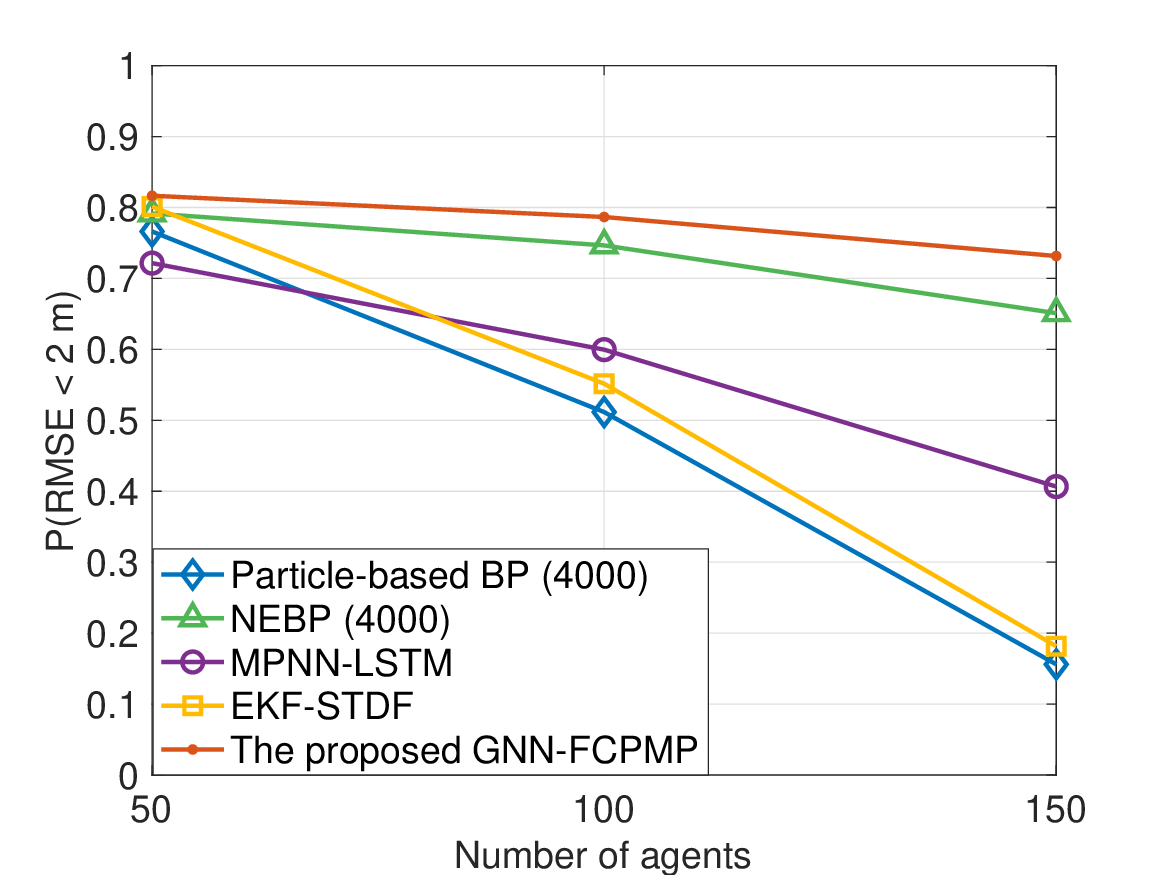}
\end{center}
\vspace*{-5mm}
\caption{Comparison of the probability that the positioning error of each agent in the network is less than 2m.}
\label{Fig_4} 
\vspace*{-5mm}
\end{figure}

In Fig.~\ref{Fig_4}, we compare the positioning performance of our GNN-FCPMP against several representative approaches, in terms of the probability \(P(\text{RMSE} \leq 2 [\text{m}])\). We set $(\sigma^t_{j \rightarrow i})^2= \left(\sigma_{i, \text {int}}^t\right)^2 = 1$, and $\ell_\text{max}$ = 10. At time slot $t$, each agent moves a distance $d_i^t\! \sim\! \mathcal{N}(2,0.1)$ in a direction $\theta_i^t\! \sim\! \mathcal{U}\left(0,2\pi\right)$. We have the following observations. Initially, the GNN enhanced algorithms, namely NEBP and GNN-FCPMP, exhibit a slight performance edge when the agent count is low. This is reasonable, because in this condition there are few short loops existing in the FG. As the number of agents increases, the performance disparity between GNN enhanced algorithms and others becomes more pronounced. This trend can be explained by two main factors: firstly, the inherent biases introduced by loops in the FG tend to propagate across the network, significantly impairing performance; secondly, the GNN enhancements applied to the loopy FG based MP effectively enhance the accuracy of the inference. Furthermore, our GNN-FCPMP consistently outperforms other alternatives due to its adept use of temporal information and the application of Chebyshev polynomials for high-precision global approximations.

\begin{figure}[t]
\vspace*{-5mm}
\begin{center}
\includegraphics[scale=0.42]{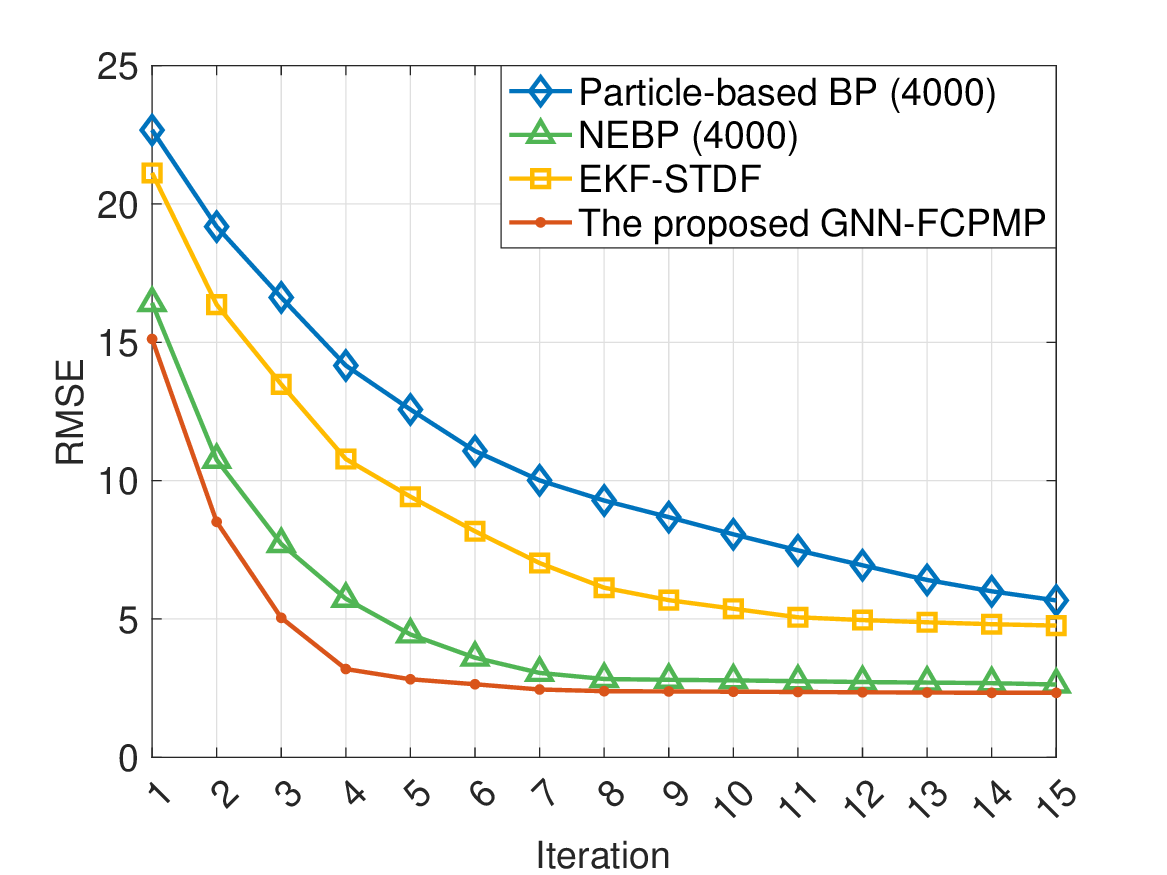}
\end{center}
\vspace*{-5mm}
\caption{The convergence of our GNN-FCPMP against NEBP, EKF-STDF and particle-based BP.}
\label{Fig_5} 
\vspace*{-5mm}
\end{figure}

Subsequently, we evaluate the convergence performance of above algorithms to gain insight. We set the number of agents to 150, $\ell_{\max} =15$, and $d_i^t\! \sim\! \mathcal{N}(3,1)$. Fig.~\ref{Fig_5} illustrates the convergence of our GNN-FCPMP against NEBP, EKF-STDF and particle-based BP. We have the following observations. Our GNN-FCPMP outshines others with its rapid convergence to the lowest RMSE, indicating swift, accurate optimal positioning and immediate stabilization post-initial iterations. In contrast, the NEBP shows commendable but slower convergence and higher RMSE, stabilizing by the 7th iteration. Particle-based BP, despite a significant number of particles (4000), lag in performance, converging at a slower rate and settling at higher RMSE values, which underscores the convergence difficulties associated with numerous short loops in the FG.

\section{Conclusion}
In dense wireless networks, conventional MP based DCP methods often struggle with inadequate accuracy of messages and convergence issues. To tackle these challenges, we have first developed a high-precision parametric message representation technique that leverages Chebyshev polynomials to approximate the nonlinear spatio-temporal messages passed on the FG. This approach significantly reduces the sensitivity of agents to the initial values of their positions at the start of the iterations. We have derived closed-form representations for all types of messages involved. Building on this parametric message representation method, we introduced a model and data-driven hybrid inference approach, the GNN-FCPMP, to avoid the intrinsic inference performance degradation caused by MP operating on loopy FG. This innovative approach is capable of fine-tuning parametric spatial messages passed on FG and obtaining more accurate closed-form expression for the \textit{a posteriori} distribution of agents’ positions. Simulation results and analyses reveal that: 1) Among all the FG-based MP DCP algorithms, our GNN-FCPMP demonstrates the best positioning performance in challenging scenarios of densely distributed mobile wireless networks; 2) Due to our globally precise parametric message representation and the refinement of messages driven by GNN, our GNN-FCPMP achieves a notably faster convergence rate compared to other MP based DCP methods.








\bibliographystyle{IEEEtran}
\bibliography{IEEEabrv,reference.bib}

\begin{thebibliography}{10}
\providecommand{\url}[1]{#1}
\csname url@samestyle\endcsname
\providecommand{\newblock}{\relax}
\providecommand{\bibinfo}[2]{#2}
\providecommand{\BIBentrySTDinterwordspacing}{\spaceskip=0pt\relax}
\providecommand{\BIBentryALTinterwordstretchfactor}{4}
\providecommand{\BIBentryALTinterwordspacing}{\spaceskip=\fontdimen2\font plus
\BIBentryALTinterwordstretchfactor\fontdimen3\font minus
  \fontdimen4\font\relax}
\providecommand{\BIBforeignlanguage}[2]{{%
\expandafter\ifx\csname l@#1\endcsname\relax
\typeout{** WARNING: IEEEtran.bst: No hyphenation pattern has been}%
\typeout{** loaded for the language `#1'. Using the pattern for}%
\typeout{** the default language instead.}%
\else
\language=\csname l@#1\endcsname
\fi
#2}}
\providecommand{\BIBdecl}{\relax}
\BIBdecl

\bibitem{Wymeersch2009}
H.~Wymeersch, J.~Lien, and M.~Z. Win, ``Cooperative localization in wireless
  networks,'' \emph{Proc. IEEE}, vol.~97, no.~2, pp. 427--450, Feb. 2009.

\bibitem{Lv_2016}
T.~Lv, H.~Gao, X.~Li, S.~Yang, and L.~Hanzo, ``Space-time hierarchical-graph
  based cooperative localization in wireless sensor networks,'' \emph{IEEE
  Transactions on Signal Processing}, vol.~64, no.~2, pp. 322--334, Jan. 2016.

\bibitem{cao2023spatial}
Y.~Cao, S.~Yang, Z.~Feng, L.~Wang, and L.~Hanzo, ``Distributed spatio-temporal
  information based cooperative 3{D} positioning in {GNSS}-denied
  environments,'' \emph{IEEE Trans. Veh. Technol.}, vol.~72, no.~1, pp. 1285 --
  1290, Jan. 2023.

\bibitem{cao2022geo}
Y.~Cao, S.~Yang, and Z.~Feng, ``Geo-spatio-temporal information based 3{D}
  cooperative positioning in {LOS/NLOS} mixed environments,'' in \emph{Proc.
  IEEE Glob. Commun. Conf. (GLOBECOM)}, Rio de Janeiro, Brazil, Dec. 2022, pp.
  5637--5642.

\bibitem{weiss1999correctness}
Y.~Weiss and W.~Freeman, ``Correctness of belief propagation in gaussian
  graphical models of arbitrary topology,'' in \emph{Proc. NeurIPS}, Denver,
  CO, USA, Nov. 1999, pp. 673--679.

\bibitem{NEBP2021}
V.~G. Satorras and M.~Welling, ``Neural enhanced belief propagation on factor
  graphs,'' in \emph{Proc. AISTATS}, San Diego, California, USA, Apr. 2021, pp.
  685--693.

\bibitem{SPAWN-comparison}
J.~Lien, U.~J. Ferner, W.~Srichavengsup, H.~Wymeersch, and M.~Z. Win, ``A
  comparison of parametric and sample-based message representation in
  cooperative localization,'' \emph{Int. Journal of Navigation and
  Observation}, vol. 2012, 10, 2012.

\bibitem{NEBP-CL2021}
M.~Liang and F.~Meyer, ``Neural enhanced belief propagation for cooperative
  localization,'' in \emph{Proc. IEEE Stat. Signal Process. Workshop (SSP)},
  Rio de Janeiro, Brazil, Jul. 2021, pp. 326--330.

\bibitem{SPATE}
N.~Wu, B.~Li, H.~Wang, C.~Xing, and J.~Kuang, ``Distributed cooperative
  localization based on {G}aussian message passing on factor graph in wireless
  networks,'' \emph{Science China Information Sciences}, vol.~58, pp. 1--15,
  2015.

\bibitem{cao2023cl}
Y.~Cao, S.~Yang, X.~Ma, and Z.~Feng, ``Cooperative positioning for sparsely
  distributed high-mobility wireless networks with {EKF} based spatio-temporal
  data fusion,'' \emph{IEEE Communications Letters}, vol.~27, no.~9, pp.
  2343--2347, Sep. 2023.

\bibitem{kirkley2021belief}
A.~Kirkley, G.~T. Cantwell, and M.~Newman, ``Belief propagation for networks
  with loops,'' \emph{Science Advances}, vol.~7, no.~17, pp. 1--9, Apr. 2021.

\bibitem{cantwell2019message}
G.~T. Cantwell and M.~E. Newman, ``Message passing on networks with loops,''
  \emph{Proc. Natl. Acad. Sci.}, vol. 116, no.~47, pp. 23\,398--23\,403, Nov.
  2019.

\bibitem{MPNN2017}
J.~Gilmer, S.~S. Schoenholz, P.~F. Riley, O.~Vinyals, and G.~E. Dahl, ``Neural
  message passing for quantum chemistry,'' in \emph{Proc. ICML}, Sydney,
  Australia, Aug. 2017, pp. 1263--1272.

\bibitem{MPNN-LSTM}
B.~C. Tedeschini, M.~Brambilla, and M.~Nicoli, ``Message passing neural network
  versus message passing algorithm for cooperative positioning,'' \emph{IEEE
  Trans. Cogn. Commun. Netw.}, vol.~9, no.~6, pp. 1666--1676, Dec. 2023.

\bibitem{li2019convergence}
B.~Li and Y.-C. Wu, ``Convergence analysis of {G}aussian belief propagation
  under high-order factorization and asynchronous scheduling,'' \emph{IEEE
  Trans. Signal Process.}, vol.~67, no.~11, pp. 2884--2897, Jun. 2019.

\end{thebibliography}


\end{document}